\documentclass[reprint,amsmath,amssymb,aps,prl,showpacs]{revtex4-1}

\usepackage{epsfig}
\usepackage{epstopdf}
\usepackage{graphicx}
\usepackage{natbib}

\begin{document}

\title{Demonstration of fold and cusp catastrophes in an atomic cloud reflected from an optical barrier in the presence of
   gravity}

\author{Serge Rosenblum}
\author{Orel Bechler}
\author{Itay Shomroni}
\author{Roy Kaner}
\author{Talya Arusi-Parpar}
\author{Oren Raz}
\author{Barak Dayan}
\email{barak.dayan@weizmann.ac.il}
\affiliation{Department of Chemical Physics, Weizmann Institute of Science}

\begin{abstract}
We experimentally demonstrate first-order (fold) and second-order (cusp) catastrophes in the density of an atomic cloud reflected from an optical barrier in the presence of gravity, and show their corresponding universal asymptotic behavior.
The cusp point enables robust, field-free refocusing of an expanding atomic cloud with a wide velocity distribution. Specifically, the density attained at the cusp point in our experiment reached 65\% of the peak density of the atoms in the trap prior to their release. We thereby add caustics to the various phenomena with parallels in optics that can be harnessed for manipulation of cold atoms. The structural stability of catastrophes provides inherent robustness against variations in the system's dynamics and initial conditions, making them suitable for manipulation of atoms under imperfect conditions and limited controllability.\\
\end{abstract}


\maketitle

From the formation of galaxy clusters in the early universe to the optical caustics
at the bottom of a pool, catastrophes play an important role in nature~\cite{berry1976,arnold1982,stewart1981,thompson1981}.
The sudden changes in the behavior of every system around a catastrophe obey the same universal laws described by singularity theory~\cite{whitney1955,thom1989,saunders1980,gilmore1981}. Catastrophes appear in nature in the common situation in which the observed behavior of a system reflects some nontrivial mapping of the parameters that govern it.
Thus, even if the phase-space distribution of the system is smooth, drastic shifts in its apparent behavior arise in places where the gradient of this mapping is zero.
Intuitively, such a mapping resembles ironing a wrinkled shirt, in which the fabric is forced onto a flat plane, leaving fold marks where the surface of the shirt was steeply curved.
Unsurprisingly, the lowest-order catastrophe is indeed called a fold catastrophe, occurring when the gradient of the mapping as a function of a single control parameter vanishes. When another control parameter is added, a higher-order catastrophe can arise, creating a `cusp' at the point where two fold lines meet~\cite{kravtsov1983}.
Catastrophes are structurally stable exactly in the same sense that slight variations in the positioning of the shirt on the ironing board may cause the fold lines to move, but not to vanish~\cite{thom1989}.
In contrast, the focusing of rays by a lens is an unstable singularity, which `dissolves' as a result of any perturbation in the lens shape or any deviation from the correct imaging conditions.
In optics, catastrophes are more widely known as caustics~\cite{berry1976,berry1980,siviloglou2007,greenfield2011,ellenbogen2009,Dolev@Arie_PRL_2012}, and their structural stability is reflected by their abundance: one does not have to carefully tune the position of his cup to observe a bright cusp feature at its bottom, and curved lines of light nearly always appear at the bottom of a swimming pool. This robustness makes catastrophes potentially useful for the manipulation of systems with limited controllability or under imperfect conditions, with examples ranging from particle manipulation with Airy beams~\cite{baumgartl2008} to focusing of electron flow~ \cite{cheianov2007} and high-harmonic generation~\cite{raz2012}.

In this work we harness catastrophes in the field of atom optics~\cite{cohen2011advances}, adding caustics to the variety of optical concepts that have already been implemented in this field by using light forces and magnetic potentials to form atomic mirrors, waveguides and lenses~\cite{adams1994,arndt1996,cohen1998manipulating}. Specifically, we demonstrate fold and cusp density catastrophes, utilizing the latter to achieve longitudinal focusing of a freely expanding atomic cloud with a large longitudinal velocity spread.
Interestingly, similar density singularities were previously studied in a completely different physical context: the large-scale structure of the universe. The evolution of an initially smooth universe under its own gravitational potential generates density singularities, known as Zeldovich pancakes~\cite{shandarin1989}, which played a role in the generation of galaxy clusters~\cite{arnold1982}. In atomic systems, caustics have been theoretically studied in a variety of configurations~\cite{Berry_JPA_1982,berry1999,ODell_JPA_2001,ODell_PRL_2012,Chalker@Shapiro_PRA_2009}, and experimentally observed with cold atoms trapped in a magnetic waveguide~\cite{Rooijakkers@Prentiss_PRA_2003} and with Bose-Einstein condensates in an optical lattice~\cite{Huckans@Porto_PRA_2009}.

\begin{figure}[t!]
\begin{center}
\includegraphics[width=1\linewidth]{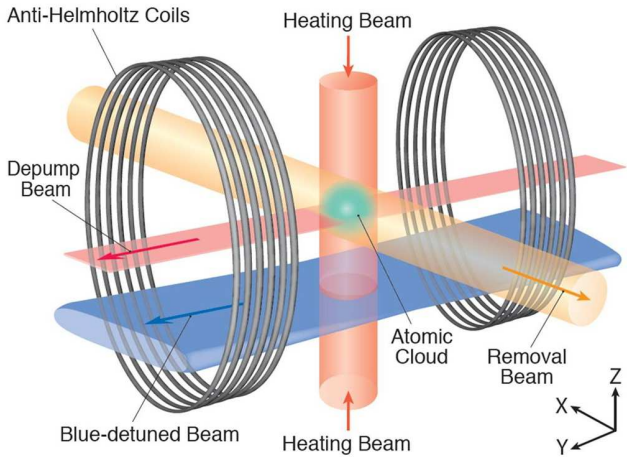}
\caption{\label{fig:experiment} Sketch of the experimental setup, showing a cloud of $\sim10^8$ $^{87}$Rb atoms at $10\,\mathrm{\mu K}$ and the optical fields that were used to tailor its initial density distribution immediately after turning off the trap (see Supplemental Material). The atoms drop due to gravity and bounce off a `floor' created by a blue-detuned beam that generates a repelling potential.\vspace{-0.9cm}}
\label{experiment}
\end{center}
\end{figure}

\begin{figure}[b]
\includegraphics[width=1\linewidth]{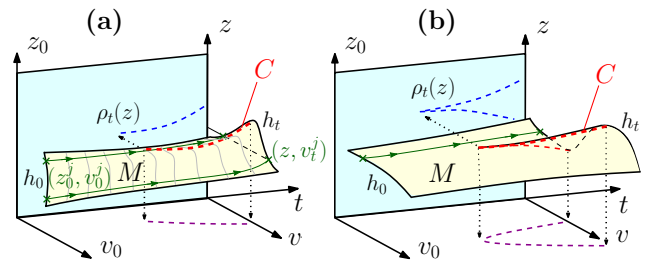}
\caption{\label{fig:Draw} Simplified illustration of two Lagrangian manifolds $M$ spanned in the same system by the phase space flow of atoms starting with different sets of initial conditions ($h_0$). The critical set $C$ (red) is a smooth curve on $M$, defining the singularity points in the projection of $M$ on the $\{z,t\}$ measurement plane. In  \textrm{(a)} the projection of $C$ on the measurement plane forms a fold line (blue). In \textrm{(b)} $C$ exhibits a parabolic turning point, as evident from its projection on the $\{v,t\}$ plane (purple). Accordingly, its projection on $\{z,t\}$ forms two fold lines that meet at a cusp point (blue). The green arrows show examples of individual atom trajectories that end at time $t$ at the same $z$, with the one drawn in \textrm{(b)} being the same as the upper trajectory in \textrm{(a)}, thus marking the intersection between the two manifolds.\vspace{0.1cm}}
\end{figure}

The case we study is the cold-atoms equivalent of a handful of marbles that are dropped to bounce on the floor. In our setup, the handful of marbles is actually a trapped cloud of $\sim10^8$ ultracold $^{87}$Rb atoms at a temperature of $10\,\mathrm{\mu K}$, and the `floor' is a blue-detuned, $40\,\mathrm{\mu m}$ thin light sheet that generates a repelling potential barrier a few millimeters below the cloud (Fig.~\ref{fig:experiment}).
At such a low temperature, once released from the trapping potential, the motion of the atoms is dominated by gravity and not by their initial velocities, and the cloud drops like a ball. If the size of the cloud is small compared to the drop height, it indeed seems to bounce off the barrier like a rigid ball (see Supplemental Material), similarly to what has been observed with evanescent-wave~\cite{aminoff1993}, magnetic~\cite{roach1995,saba1999,Arnold@Boshier_JPB_2004} and optical dipole~\cite{bongs1999} atomic mirrors.
However, since the atoms in our case are non-interacting, the rigid ball description is misleading. The accurate description of the dynamics along the vertical $z$ direction, which is the axis of interest, is therefore the propagation of an ensemble of particles that begin with some phase-space distribution of initial heights and velocities $\{z_0,v_0\}$. As each atom propagates independently, this initial phase-space distribution evolves according to Hamilton's equations. This can be considered as a time-dependent mapping between initial Lagrangian coordinates $\{z_0,v_0\}$ to Eulerian coordinates $\{z,v\}$ which describe the resulting height and position as a function of time.
As depicted in Fig.~\ref{fig:Draw}a, if the initial distribution can be assumed to lie along some one-dimensional starting curve  $h_0(z_0,v_0)$, then the evolution of this curve defines a two-dimensional smooth manifold $M$, called the Lagrangian manifold, embedded in the three-dimensional space $\left\{z,v,t\right\}$.
An imaging experiment measures the time-dependent linear density $\rho_t\left(z\right)$ of the atoms, yet ignores their velocities. It thus represents a projection of $M$ onto the two-dimensional ``measurement plane'' $\left\{z,t\right\}$. Since at each point in time $M$ may define a few single-valued velocity functions $v_t^j(z)$, the projections of the corresponding linear densities $\lambda_t^j(z)$ need to be summed:
\begin{eqnarray}
\label{rho}
\rho_t\left(z\right)dz
= \sum_j\lambda_t^j(z)\sqrt{dz^2+\left(\partial_z v_t^j(z)\:dz\right)^2 }  \: .
\end{eqnarray}
The set of points at which the gradient $\partial_z v_t^j(z)$ diverges is called the critical set of the mapping. This is the set of points at which $M$ is perpendicular to the measurement plane, leading to singularities in $\rho_t\left(z\right)$. As was shown by Whitney~\cite{whitney1955}, this critical set forms a smooth curve $C$ on the manifold $M$. The resulting types of singularities can be defined by the derivatives of the projection of $C$ on the measurement plane (the blue curves in Fig.~\ref{fig:Draw}) as a function of the position along the curve itself. When this derivative is nonzero the singularity is called a fold. A cusp point is defined when $C$ has a turning point, as depicted in Fig.~\ref{fig:Draw}b, namely when the first derivative is zero, but the second is not.
All higher-order singularities are unstable in two dimensions, in the sense that even small perturbations dissolve them into the structurally-stable fold or cusp catastrophes~\cite{whitney1955}.
As illustrated in Figs.~\ref{fig:Draw}a and~\ref{fig:Draw}b, which represent different experiments performed in the same system, it is the specific choice of initial conditions that defines which singularities may be observed in each experiment.

\begin{figure}
\begin{center}
\includegraphics[width=0.5\textwidth]{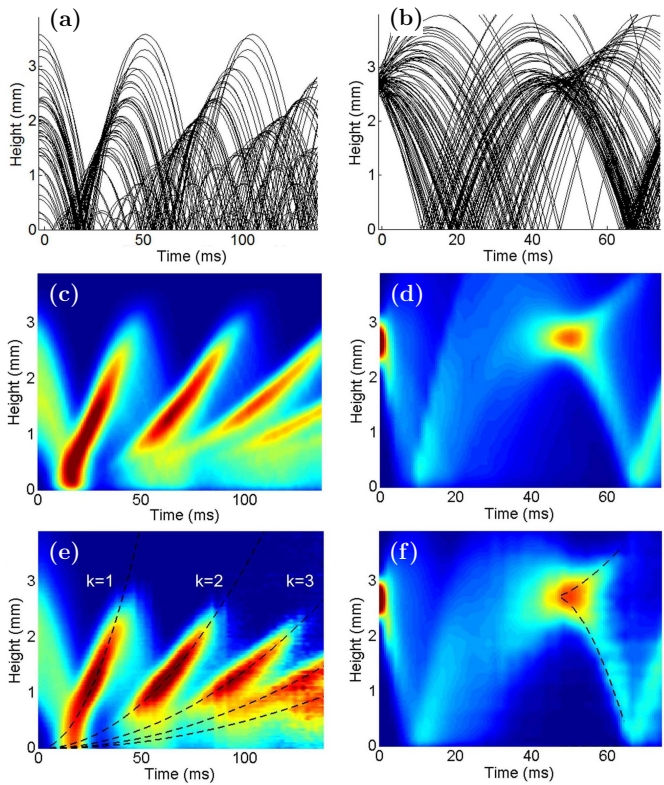}
\caption{\label{fig:Results} Calculated, simulated and experimental results. \textrm{(a)} Calculated trajectories of atoms released from a cold ($10\,\mathrm{\mu K}$) but large atomic cloud (standard deviation of $\sim 1\,\mathrm{mm}$, compared to release height of $\sim2\,\mathrm{mm}$). \textrm{(b)} Calculated trajectories of atoms released from a hot ($\sim400\,\mathrm{\mu K}$) pancake-shaped atomic cloud (standard deviation of $\sim 0.1\,\mathrm{mm}$, compared to release height of $\sim2.8\,\mathrm{mm}$), demonstrating refocusing of the cloud after $\sim 48\,\mathrm{ms}$, followed by splitting into two high-density peaks, one moving upwards and the other downwards. \textrm{(c),(d)} Simulated results taking into account the experimental initial conditions and the temporal and spatial resolution of the imaging setup, clearly showing the emergence of fold lines in (c), and a cusp splitting into two folds in (d). \textrm{(e),(f)} The  experimentally measured vertical density distributions as a function of time, together with the analytically derived location of the catastrophe lines (dashed). The density attained at the cusp point is $\sim 65\%$ of the initial density of the expanding atomic cloud at $t=0$.}
\label{results}
\end{center}
\end{figure}

\begin{figure}[b]
\includegraphics[width=0.505\textwidth]{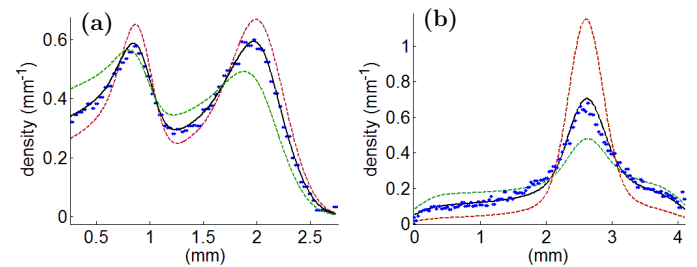}
\caption{\label{fig:Fits} \textrm{(a)} For fold type singularities, the measured density distribution at a given time (dots) is in excellent agreement with the theoretical distribution (solid line) using a divergence rate of $\alpha=1/2$. In comparison, divergence rates of $\alpha=1/3$ (green dashed line) and of $\alpha=2/3$ (red dashed line) are inconsistent with the measured results. \textrm{(b)} For a cusp type singularity, the measured density distribution (dots) agrees with the theoretical distribution (solid line) using $\alpha=2/3$, and is inconsistent with divergence rates of $\alpha=1/2$ (green dashed line) and of $\alpha=1$ (red dashed line).\vspace{-0.9cm}}
\end{figure}

In our experiment, in order to demonstrate fold catastrophes, we chose to span the manifold in which $h_0$ is close to a vertical line, i.e. a very small spread in $v_0$, and a large spread in $z_0$.
To realize this manifold, we let a cold ($10\,\mathrm{\mu K}$) but large atomic cloud (standard deviation of $\sim 1\,\mathrm{mm}$ in $z_0$) drop from a height of $\sim2\,\mathrm{mm}$. The height distribution of the atoms as a function of time, extracted from a sequence of fluorescence images (see Supplemental Material), differs drastically from the behavior of a rigid ball (Fig.~\ref{fig:Results}e).
Surprisingly, once released, the smooth distribution of the falling atomic cloud gives rise to a series of density waves moving upwards from the barrier below.
Similar behavior was previously observed in an experiment for atom accumulation using an optical ``trap door''~\cite{davies2000}.
The underlying dynamics behind these waves are illustrated in Fig.~\ref{fig:Results}a, which presents the trajectories of individual atoms assuming the same initial height distribution and velocity spread as in the experiment. Specifically, atoms that begin at zero velocity one slightly above the other switch places at the instant of the bounce, since the lower atom arrives to the barrier first. However, conservation of energy implies that the higher atom must eventually cross the lower one to end back on top. This intersection of the trajectories of atoms from slightly different $z_0$ as they approach their peak height gives rise to density singularities. Since this coalescence happens later for atoms that began higher, these density peaks seem to rise constantly with time, even though each of the individual atoms within the peaks follows the regular parabolic rise and fall.
The critical set for this manifold was derived by neglecting the initial velocity spread of the atoms and rewriting Eq.~\eqref{rho} as a function of the initial linear density distribution $\lambda_0(z_0)$ along $h_0$:
\begin{equation}
\rho_t\left(z\right) = \sum_j \lambda_0\left(z_{0,t}^j\right) \left|\partial_z z_{0,t}^j\right|\: ,
\end{equation}
with ${z_{0,t}^j(z)}$ being the set of initial positions that are mapped at time $t$ to $z$.
The solution of this expression leads to a set of rising parabolic fold lines (dashed lines in Fig.~\ref{fig:Results}e), one for every bounce number $k$:
\begin{eqnarray}
\label{zfold}
z_\textsl{fold}(t) = \frac{1}{2}\frac{g}{(4k^2-1)}t^2 .
\end{eqnarray}
$\text{   }$ $\text{   }$
The density peaks are composed of atoms whose trajectory is tangent to this parabola, which occurs when the atoms reach $1-(2k)^{-2}$ of their initial height. Also evident from Fig.~\ref{fig:Results}a is the fact that the density peak lines separate between regions in which there are two valid atom trajectories (for a given $k$) and regions where there are none, which is indeed a property of fold lines. Another property of catastrophes of this type, derived by Arnold, Shandarin and Zeldovich in their study on the large scale structure of the universe~\cite{arnold1982}, is asymptotic divergence of the form $\rho \propto z^{-\alpha}$, with $\alpha=1/2$ for a fold, and $\alpha=2/3$ for a cusp. Taking into consideration the measured initial parameters of the atomic cloud and the spatial resolution, the density distribution measured at the vicinity of the density peaks yields excellent fit to this behavior (Fig.~\ref{fig:Fits}a). \\$\text{   }$ $\text{   }$

Viewing the dynamics of the atomic cloud as topological mapping provides the means for identifying the conditions in which it will exhibit catastrophic behavior of higher orders as well. In particular, locating a second-order cusp point bears practical significance since, being a turning point of a fold-line, a cusp is  closer to ideal focusing, as it can create the situation in which most of the population of the atomic cloud is concentrated at the vicinity of a single point. Moreover, as described earlier, a cusp point is in fact as close as one can get to ideal focusing while remaining structurally stable (in 2D).
Generally, the search for singularities becomes more complex at higher orders, yet the knowledge of the position of the first-order fold lines can direct us in this search.
Specifically, from the solution to Hamilton's equations (see Supplemental Material) it can be shown that the projection of the critical set of the previous manifold on the $\{v,t\}$ plane forms straight lines, implying that $C$ is never perpendicular to the measurement plane.
This rules out the emergence of higher-order catastrophes with this set of initial conditions.
Thus, to demonstrate a cusp catastrophe we chose the opposite set of initial conditions: a small spread in $z_0$, and a large spread in $v_0$, corresponding to a manifold that is initially perpendicular to the previous one ($h_0$ being a horizontal line).
To create such initial conditions, we heated the cloud to $\sim400\,\mathrm{\mu K}$ using counter-propagating heating beams along the $z$-direction (see Fig.~\ref{experiment}), and then reduced its vertical spread to $0.1$ mm (see Supplemental Material).

As presented in Fig.~\ref{fig:Results}f, the atoms initially expand ballistically and the density drops drastically. However, after $\sim48\,\mathrm{ms}$ the cloud refocuses, forming a high-density peak which then splits to form two density peaks, one moving upwards and one downwards.
The trajectories plotted in Fig.~\ref{fig:Results}b reveal the underlying mechanism of this effect, which is indeed a cusp singularity that evolves into two fold lines.  The analytic expression for the critical set $C$ this time was derived by assuming negligible initial spatial spread and solving:
\begin{equation}
\rho_t\left(z\right) =  \sum_j \lambda_0\left(v_{0,t}^j\right) \left|\partial_z v_{0,t}^j\right| .
\end{equation}
The resulting dynamics indeed show divergence of the density in the form of a cusp point that occurs after each bounce, as the atoms that began at rest reach back their original (and maximal) height, which then evolves into two fold lines~\cite{Catastrophe_another_fold} (dashed lines in Fig.~\ref{fig:Results}f).  The top fold line is composed of atoms that began with positive initial velocity, and the other is composed of atoms that began with negative velocity. The turning point between positive and negative initial velocities generates the observed focusing at the cusp point. The divergence rate obtained from the measured density distribution is in excellent agreement with the theoretical value of $2/3$ (Fig.~\ref{fig:Fits}b). The measured density at the cusp point reached $\sim65\%$ of the initial density of the cloud, immediately after its release from the trap. Note that such focusing in the longitudinal direction cannot be achieved with lens-like potentials, as such cannot distinguish between fast and slow atoms that travel along the same path. In essence, this spatial focusing of atoms with a wide velocity distribution can be viewed as complementary to the method of $\delta$-kick cooling, which results in atoms that are cold, yet spread spatially~\cite{Hubert@Nelson_PRL_1997,Myrskog@Steinberg_PRA_2000,Jendrzejewski@Josse_PRL_2012}.\\ $\text{   }$ $\text{   }$
Analyzing the density evolution of cold atoms in terms of catastrophe theory does not only place this system in its appropriate, universal context; it also accurately reveals the rules that govern its dynamics. The singularities demonstrated here are practically insensitive to the parameters of the reflecting potential in the same way that slight stretching or twisting of a folded sheet does not eliminate the fold. The robustness to the initial conditions stems from the fact that the critical set is completely defined by the shape of $h_0$, regardless of the exact density distribution along this curve. In other words, the dashed lines in Figs.~\ref{fig:Results}e and \ref{fig:Results}f, are already there once the manifold is defined, and the specific initial density distribution only `populates' these lines. For example, had we started the cusp experiment with only positive velocities, we would have observed only the upward-going fold evolving from the cusp point, yet the location of this fold and the cusp point would have remained unchanged. Finally, the density singularities we demonstrated evolve out of free propagation, and so enable focusing away from any trapping (and perturbing) fields. Integrating caustics into the toolbox of atom optics therefore provides an alternative, versatile mechanism, particularly relevant when the analogy of the atomic ensemble to optical beams fails.


\clearpage

\onecolumngrid

\section{Supplemental Material}
\noindent\emph{Experimental details}

\begin{figure*}[h!]
\begin{center}
\includegraphics[width=0.7\linewidth]{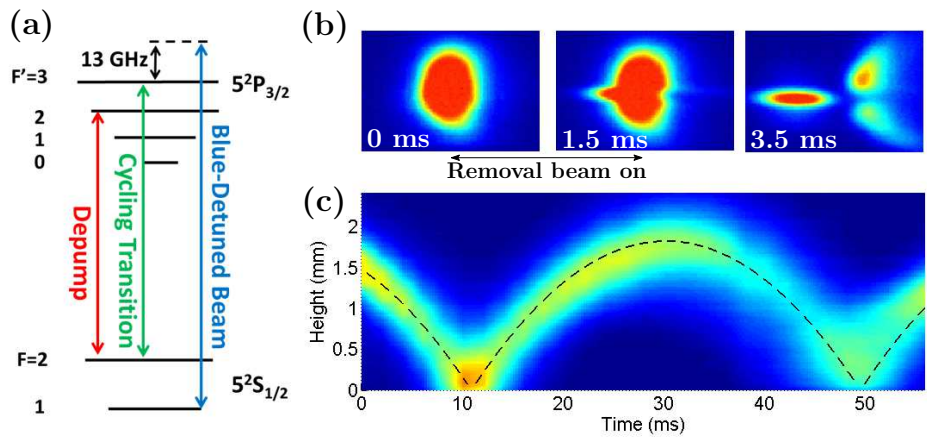}
\caption{\label{fig:FigSupp}  \textrm{(a)} Energy levels of the $^{87}$Rb $D_2$ transition. The heating beam, used for tuning the cloud's vertical velocity distribution, as well as the removal beam, are resonant with the cycling transition.
\textrm{(b)} The initially round atomic cloud (left fluorescence image) can be brought to a thin pancake shape by irradiating only its middle section with the depump beam, which pumps the atoms to the $F=1$ ground state. The removal beam is then used to blow away the rest of the atoms that remained in $F=2$ (middle and right images).
\textrm{(c)} Measurement of the atomic density along the $z$-axis as a function of time for a bouncing atomic cloud with small initial spatial spread (standard deviation $\sim0.1\,\mathrm{mm}$), and small vertical velocity spread (standard deviation $\sim0.03\,\mathrm{m/s}$). The cloud follows the same parabolic trajectory expected of a rigid ball (dashed line).\vspace{-0.9cm}}
\label{experiment}
\end{center}
\end{figure*}

The atoms were collected using a Magneto-Optical trap, and were depumped to the $F=1$ hyperfine state prior to their release (Fig.~\ref{fig:FigSupp}a).
The repelling light sheet was a $\sim 400\,\mathrm{mW}$ elliptical beam with vertical waist of $40\,\mathrm{\mu m}$ and horizontal waist of $1\,\mathrm{cm}$ (full width, $1/e^2$), and was blue-detuned from the $F=1 \rightarrow F'=3$ resonance by $13\,\mathrm{GHz}$ in the fold experiment, and by $5\,\mathrm{GHz}$ in the cusp experiment.

Reducing the initial vertical distribution of the cloud was accomplished by using an elliptical beam to pump the atoms from $F=2$ to $F=1$ only at a thin slice at the center of the atomic cloud, and then shining a removal beam resonant with the $F=2 \rightarrow F'=3$ cycling transition to blow away the rest of the atoms (Fig.~\ref{fig:FigSupp}b). If the temperature of the cloud is kept low enough, the motion of the atoms is dominated by gravity and not by their initial velocities, and the cloud bounces like a rigid ball (Fig.~\ref{fig:FigSupp}c), tracing the expected parabolic path determined by Earth's gravitational acceleration $g$.\\\\
\noindent\emph{Imaging setup}

Fluorescence images of the falling atoms (as in Fig.~\ref{fig:FigSupp}b) were captured by a CCD camera using $1\,\mathrm{ms}$ long pulses of resonant light. Since this process scatters the atoms, every measurement required a new loading and dropping cycle, with varying fall durations before the imaging. To present the vertical density as a function of time (as in Fig.~\ref{fig:FigSupp}c) we integrated consecutive fluorescence images horizontally, and cascaded the resulting column vectors.\\\\

\newpage
\noindent\emph{Simulations}

The simulations were performed by propagating the position of $10^4$ atoms with given initial position and velocity distributions, and then blurring the image with a convolution kernel whose width equals the spatial resolution.
The solution of Hamilton's equations for an atom that accelerates downward at $g$ and bounces from a perfect barrier, given the bounce number $k$ is:

\begin{eqnarray}
z(t) &=& (1-4k^2)z_0 + v_0 t - \frac{g t^2}{2}\label{zt}\\
   &&+ 2k(t-v_0/g) \sqrt{v_0^2 + 2z_0 g}- \frac{2 v_0^2}{g}k^2\nonumber\label{vt}\\
v(t) &=& v_0 - g t + 2k\sqrt{v_0^2 + 2z_0 g}.
\end{eqnarray}\label{density}
While the bounce number $k$ can be trivially calculated from $z_0,v_0,t$, it is more easily defined as the only $k$ for which $z(z_0,v_0,t,k)$ is positive.


\begin{thebibliography}{10}

\expandafter\ifx\csname url\endcsname\relax
  \def\url#1{\texttt{#1}}\fi
\expandafter\ifx\csname urlprefix\endcsname\relax\def\urlprefix{URL }\fi
\providecommand{\bibinfo}[2]{#2}
\providecommand{\eprint}[2][]{\url{#2}}

\bibitem{berry1976}
\bibinfo{author}{Berry, M.}
\newblock \bibinfo{title}{Waves and Thom's theorem}.
\newblock \emph{\bibinfo{journal}{Advances in Physics}}
  \textbf{\bibinfo{volume}{25}}, \bibinfo{pages}{1--26} (\bibinfo{year}{1976}).

\bibitem{stewart1981}
\bibinfo{author}{Stewart, I.}
\newblock \bibinfo{title}{Applications of catastrophe theory to the physical
  sciences}.
\newblock \emph{\bibinfo{journal}{Physica D: Nonlinear Phenomena}}
  \textbf{\bibinfo{volume}{2}}, \bibinfo{pages}{245 -- 305}
  (\bibinfo{year}{1981}).

\bibitem{thompson1981}
\bibinfo{author}{Thompson, J.}
\newblock \emph{\bibinfo{title}{Instabilities and catastrophes in science and
  engineering}} (\bibinfo{publisher}{John Wiley \& Sons Ltd},
  \bibinfo{year}{1981}).

\bibitem{arnold1982}
\bibinfo{author}{Arnold, V.~I.}, \bibinfo{author}{Shandarin, S.~F.} \&
  \bibinfo{author}{Zeldovich, Y.~B.}
\newblock \bibinfo{title}{The large scale structure of the universe I.
  general properties. one-and two-dimensional models}.
\newblock \emph{\bibinfo{journal}{Geophysical \& Astrophysical Fluid Dynamics}}
  \textbf{\bibinfo{volume}{20}}, \bibinfo{pages}{111--130}
  (\bibinfo{year}{1982}).

\bibitem{saunders1980}
\bibinfo{author}{Saunders, P.~T.}
\newblock \emph{\bibinfo{title}{An Introduction to Catastrophe Theory}}
  (\bibinfo{publisher}{Cambridge University Press}, \bibinfo{year}{1980}).

\bibitem{whitney1955}
\bibinfo{author}{Whitney, H.}
\newblock \bibinfo{title}{On singularities of mappings of Euclidean
  spaces. I. mappings of the plane into the plane}.
\newblock \emph{\bibinfo{journal}{The Annals of Mathematics}}
  \textbf{\bibinfo{volume}{62}}, \bibinfo{pages}{pp. 374--410}
  (\bibinfo{year}{1955}).

\bibitem{gilmore1981}
\bibinfo{author}{Gilmore, R.}
\newblock \emph{\bibinfo{title}{Catastrophe Theory for Scientists and
  Engineers}} (\bibinfo{publisher}{Dover, New York}, \bibinfo{year}{1981}).

\bibitem{thom1989}
\bibinfo{author}{Thom, R.}
\newblock \emph{\bibinfo{title}{Structural stability and morphogenesis}}
  (\bibinfo{publisher}{Addison Wesley Publishing Company},
  \bibinfo{year}{1989}).

\bibitem{kravtsov1983}
\bibinfo{author}{Kravtsov, Y.~A.} \& \bibinfo{author}{Orlov, Y.~I.}
\newblock \bibinfo{title}{Caustics, catastrophes, and wave fields}.
\newblock \emph{\bibinfo{journal}{Soviet Physics Uspekhi}}
  \textbf{\bibinfo{volume}{26}}, \bibinfo{pages}{1038} (\bibinfo{year}{1983}).

\bibitem{berry1980}
\bibinfo{author}{Berry, M.} \& \bibinfo{author}{Upstill, C.}
\newblock \bibinfo{title}{IV catastrophe optics: Morphologies of
  caustics and their diffraction patterns}.
\newblock vol.~\bibinfo{volume}{18} of \emph{\bibinfo{series}{Progress in
  Optics}}, \bibinfo{pages}{257 -- 346} (\bibinfo{publisher}{Elsevier},
  \bibinfo{year}{1980}).

\bibitem{siviloglou2007}
\bibinfo{author}{Siviloglou, G.~A.}, \bibinfo{author}{Broky, J.},
  \bibinfo{author}{Dogariu, A.} \& \bibinfo{author}{Christodoulides, D.~N.}
\newblock \bibinfo{title}{Observation of accelerating airy beams}.
\newblock \emph{\bibinfo{journal}{Phys. Rev. Lett.}}
  \textbf{\bibinfo{volume}{99}}, \bibinfo{pages}{213901}
  (\bibinfo{year}{2007}).

\bibitem{greenfield2011}
\bibinfo{author}{Greenfield, E.}, \bibinfo{author}{Segev, M.},
  \bibinfo{author}{Walasik, W.} \& \bibinfo{author}{Raz, O.}
\newblock \bibinfo{title}{Accelerating light beams along arbitrary convex
  trajectories}.
\newblock \emph{\bibinfo{journal}{Phys. Rev. Lett.}}
  \textbf{\bibinfo{volume}{106}}, \bibinfo{pages}{213902}
  (\bibinfo{year}{2011}).

\bibitem{ellenbogen2009}
\bibinfo{author}{Ellenbogen, T.}, \bibinfo{author}{Voloch-Bloch, N.},
  \bibinfo{author}{Ganany-Padowicz, A.} \& \bibinfo{author}{Arie, A.}
\newblock \bibinfo{title}{Nonlinear generation and manipulation of
  Airy beams}.
\newblock \emph{\bibinfo{journal}{Nature Photon.}}
  \textbf{\bibinfo{volume}{3}}, \bibinfo{pages}{395--398}
  (\bibinfo{year}{2009}).

\bibitem{Dolev@Arie_PRL_2012}
\bibinfo{author}{Dolev, I.}, \bibinfo{author}{Kaminer, I.},
  \bibinfo{author}{Shapira, A.}, \bibinfo{author}{Segev, M.} \&
  \bibinfo{author}{Arie, A.}
\newblock \bibinfo{title}{Experimental observation of self-accelerating beams
  in quadratic nonlinear media}.
\newblock \emph{\bibinfo{journal}{Phys. Rev. Lett.}}
  \textbf{\bibinfo{volume}{108}}, \bibinfo{pages}{113903}
  (\bibinfo{year}{2012}).

\bibitem{baumgartl2008}
\bibinfo{author}{Baumgartl, J.}, \bibinfo{author}{Mazilu, M.} \&
  \bibinfo{author}{Dholakia, K.}
\newblock \bibinfo{title}{Optically mediated particle clearing using
  Airy wavepackets}.
\newblock \emph{\bibinfo{journal}{Nature Photon.}}
  \textbf{\bibinfo{volume}{2}}, \bibinfo{pages}{675--678}
  (\bibinfo{year}{2008}).

\bibitem{cheianov2007}
\bibinfo{author}{Cheianov, V.~V.}, \bibinfo{author}{Fal'ko, V.} \&
  \bibinfo{author}{Altshuler, B.~L.}
\newblock \bibinfo{title}{The focusing of electron flow and a Veselago
  lens in graphene p-n junctions}.
\newblock \emph{\bibinfo{journal}{Science}} \textbf{\bibinfo{volume}{315}},
  \bibinfo{pages}{1252--1255} (\bibinfo{year}{2007}).

\bibitem{raz2012}
\bibinfo{author}{Raz, O.}, \bibinfo{author}{Pedatzur, O.},
  \bibinfo{author}{Bruner, B.~D.} \& \bibinfo{author}{Dudovich, N.}
\newblock \bibinfo{title}{Spectral caustics in attosecond science}.
\newblock \emph{\bibinfo{journal}{Nature Photon.}}
  \textbf{\bibinfo{volume}{6}}, \bibinfo{pages}{170--173}
  (\bibinfo{year}{2012}).

\bibitem{cohen2011advances}
\bibinfo{author}{Cohen-Tannoudji, C.} \& \bibinfo{author}{Gu{\'e}ry-Odelin, D.}
\newblock \emph{\bibinfo{title}{Advances in atomic physics}}
  (\bibinfo{publisher}{World Scientific}, \bibinfo{year}{2011}).

\bibitem{adams1994}
\bibinfo{author}{Adams, C.~S.}
\newblock \bibinfo{title}{Atom optics}.
\newblock \emph{\bibinfo{journal}{Contemporary Physics}}
  \textbf{\bibinfo{volume}{35}}, \bibinfo{pages}{1--19} (\bibinfo{year}{1994}).

\bibitem{arndt1996}
\bibinfo{author}{Arndt, M.}, \bibinfo{author}{Szriftgiser, P.},
  \bibinfo{author}{Dalibard, J.} \& \bibinfo{author}{Steane, A.~M.}
\newblock \bibinfo{title}{Atom optics in the time domain}.
\newblock \emph{\bibinfo{journal}{Phys. Rev. A}} \textbf{\bibinfo{volume}{53}},
  \bibinfo{pages}{3369--3378} (\bibinfo{year}{1996}).

\bibitem{cohen1998manipulating}
\bibinfo{author}{Cohen-Tannoudji, C.}
\newblock \bibinfo{title}{Manipulating atoms with photons}.
\newblock \emph{\bibinfo{journal}{Physica Scripta}}
  \textbf{\bibinfo{volume}{1998}}, \bibinfo{pages}{33} (\bibinfo{year}{1998}).

\bibitem{shandarin1989}
\bibinfo{author}{Shandarin, S.~F.} \& \bibinfo{author}{Zeldovich, Y.~B.}
\newblock \bibinfo{title}{The large-scale structure of the universe:
  Turbulence, intermittency, structures in a self-gravitating medium}.
\newblock \emph{\bibinfo{journal}{Rev. Mod. Phys.}}
  \textbf{\bibinfo{volume}{61}}, \bibinfo{pages}{185--220}
  (\bibinfo{year}{1989}).

\bibitem{Berry_JPA_1982}
\bibinfo{author}{Berry, M.~V.}
\newblock \bibinfo{title}{Wavelength-independent fringe spacing in rainbows
  from falling neutrons}.
\newblock \emph{\bibinfo{journal}{Journal of Physics A: Mathematical and
  General}} \textbf{\bibinfo{volume}{15}}, \bibinfo{pages}{L385}.

\bibitem{berry1999}
\bibinfo{author}{Berry, M.} \& \bibinfo{author}{O'Dell, D.}
\newblock \bibinfo{title}{Ergodicity in wave-wave diffraction}.
\newblock \emph{\bibinfo{journal}{Journal of Physics A: Mathematical and
  General}} \textbf{\bibinfo{volume}{32}}, \bibinfo{pages}{3571}
  (\bibinfo{year}{1999}).

\bibitem{ODell_JPA_2001}
\bibinfo{author}{O'Dell, D. H.~J.}
\newblock \bibinfo{title}{Dynamical diffraction in sinusoidal potentials:
  uniform approximations for mathieu functions}.
\newblock \emph{\bibinfo{journal}{Journal of Physics A: Mathematical and
  General}} \textbf{\bibinfo{volume}{34}}, \bibinfo{pages}{3897}
  (\bibinfo{year}{2001}).

\bibitem{ODell_PRL_2012}
\bibinfo{author}{O'Dell, D. H.~J.}
\newblock \bibinfo{title}{Quantum catastrophes and ergodicity in the dynamics
  of bosonic josephson junctions}.
\newblock \emph{\bibinfo{journal}{Phys. Rev. Lett.}}
  \textbf{\bibinfo{volume}{109}}, \bibinfo{pages}{150406}
  (\bibinfo{year}{2012}).

\bibitem{Chalker@Shapiro_PRA_2009}
\bibinfo{author}{Chalker, J.~T.} \& \bibinfo{author}{Shapiro, B.}
\newblock \bibinfo{title}{Caustic formation in expanding condensates of cold
  atoms}.
\newblock \emph{\bibinfo{journal}{Phys. Rev. A}} \textbf{\bibinfo{volume}{80}},
  \bibinfo{pages}{013603} (\bibinfo{year}{2009}).

\bibitem{Rooijakkers@Prentiss_PRA_2003}
\bibinfo{author}{Rooijakkers, W.} \emph{et~al.}
\newblock \bibinfo{title}{Observation of caustics in the trajectories of cold
  atoms in a linear magnetic potential}.
\newblock \emph{\bibinfo{journal}{Phys. Rev. A}} \textbf{\bibinfo{volume}{68}},
  \bibinfo{pages}{063412} (\bibinfo{year}{2003}).

\bibitem{Huckans@Porto_PRA_2009}
\bibinfo{author}{Huckans, J.~H.}, \bibinfo{author}{Spielman, I.~B.},
  \bibinfo{author}{Tolra, B.~L.}, \bibinfo{author}{Phillips, W.~D.} \&
  \bibinfo{author}{Porto., J.~V.}
\newblock \bibinfo{title}{Quantum and classical dynamics of a Bose-Einstein
  condensate in a large-period optical lattice}.
\newblock \emph{\bibinfo{journal}{Phys. Rev. A}} \textbf{\bibinfo{volume}{80}},
  \bibinfo{pages}{043609} (\bibinfo{year}{2009}).

\bibitem{aminoff1993}
\bibinfo{author}{Aminoff, C.~G.} \emph{et~al.}
\newblock \bibinfo{title}{Cesium atoms bouncing in a stable gravitational
  cavity}.
\newblock \emph{\bibinfo{journal}{Phys. Rev. Lett.}}
  \textbf{\bibinfo{volume}{71}}, \bibinfo{pages}{3083--3086}
  (\bibinfo{year}{1993}).

\bibitem{roach1995}
\bibinfo{author}{Roach, T.~M.} \emph{et~al.}
\newblock \bibinfo{title}{Realization of a magnetic mirror for cold atoms}.
\newblock \emph{\bibinfo{journal}{Phys. Rev. Lett.}}
  \textbf{\bibinfo{volume}{75}}, \bibinfo{pages}{629--632}
  (\bibinfo{year}{1995}).

\bibitem{saba1999}
\bibinfo{author}{Saba, C.~V.} \emph{et~al.}
\newblock \bibinfo{title}{Reconstruction of a cold atom cloud by magnetic
  focusing}.
\newblock \emph{\bibinfo{journal}{Phys. Rev. Lett.}}
  \textbf{\bibinfo{volume}{82}}, \bibinfo{pages}{468--471}
  (\bibinfo{year}{1999}).

\bibitem{Arnold@Boshier_JPB_2004}
\bibinfo{author}{Arnold, A.~S.}, \bibinfo{author}{MacCormick, C.} \&
  \bibinfo{author}{Boshier, M.~G.}
\newblock \bibinfo{title}{Diffraction-limited focusing of Bose-Einstein
  condensates}.
\newblock \emph{\bibinfo{journal}{Journal of Physics B: Atomic, Molecular and
  Optical Physics}} \textbf{\bibinfo{volume}{37}}, \bibinfo{pages}{485}.

\bibitem{bongs1999}
\bibinfo{author}{Bongs, K.} \emph{et~al.}
\newblock \bibinfo{title}{Coherent evolution of bouncing Bose-Einstein
  condensates}.
\newblock \emph{\bibinfo{journal}{Phys. Rev. Lett.}}
  \textbf{\bibinfo{volume}{83}}, \bibinfo{pages}{3577--3580}
  (\bibinfo{year}{1999}).

\bibitem{davies2000}
\bibinfo{author}{Davies, H.~J.}, \bibinfo{author}{Szymaniec, K.} \&
  \bibinfo{author}{Adams, C.~S.}
\newblock \bibinfo{title}{Cold-atom accumulation using an optical trap door}.
\newblock \emph{\bibinfo{journal}{Phys. Rev. A}} \textbf{\bibinfo{volume}{62}},
  \bibinfo{pages}{013412} (\bibinfo{year}{2000}).

\bibitem{Catastrophe_another_fold}
\bibinfo{note}{For all $k>1$, the reflection of the downward fold from the
  previous cusp yields another fold moving upwards.}

\bibitem{Hubert@Nelson_PRL_1997}
\bibinfo{author}{Ammann, H.} \& \bibinfo{author}{Christensen, N.}
\newblock \bibinfo{title}{Delta kick cooling: A new method for cooling atoms}.
\newblock \emph{\bibinfo{journal}{Phys. Rev. Lett.}}
  \textbf{\bibinfo{volume}{78}}, \bibinfo{pages}{2088--2091}
  (\bibinfo{year}{1997}).

\bibitem{Myrskog@Steinberg_PRA_2000}
\bibinfo{author}{Myrskog, S.~H.}, \bibinfo{author}{Fox, J.~K.},
  \bibinfo{author}{Moon, H.~S.}, \bibinfo{author}{Kim, J.~B.} \&
  \bibinfo{author}{Steinberg, A.~M.}
\newblock \bibinfo{title}{Modified ``$\delta${}-kick cooling'' using magnetic
  field gradients}.
\newblock \emph{\bibinfo{journal}{Phys. Rev. A}} \textbf{\bibinfo{volume}{61}},
  \bibinfo{pages}{053412} (\bibinfo{year}{2000}).

\bibitem{Jendrzejewski@Josse_PRL_2012}
\bibinfo{author}{Jendrzejewski, F.} \emph{et~al.}
\newblock \bibinfo{title}{Coherent backscattering of ultracold atoms}.
\newblock \emph{\bibinfo{journal}{Phys. Rev. Lett.}}
  \textbf{\bibinfo{volume}{109}}, \bibinfo{pages}{195302}
  (\bibinfo{year}{2012}).

\end{thebibliography}
\end{document}